\documentclass{edp-jp4}
\usepackage{graphicx}
\begin{document}

\title{History dependence, memory and metastability in electron glasses} 
\author{Markus M\"uller}\address{Center for Materials Theory, Serin Physics Laboratory,
    Rutgers University,\\
136 Frelinghuysen Road, Piscataway, New Jersey 08854-8019, USA}
\author{Eran Lebanon}\sameaddress{1}
\maketitle
\begin{abstract} 
We discuss the history dependence and memory effects which are observed in the out-of-equilibrium conductivity of electron glasses. The experiments can be understood by assuming that the local density of states retains a memory of the sample history.  We provide analytical arguments for the consistency of this assumption, and discuss the saturation of the memory effect with increasing gate voltage change.
This picture is bolstered by 
numerical simulations at zero temperature, 
which moreover demonstrate the incompressibility of the Coulomb glass on short timescales.
\end{abstract}
\section{Introduction}
Glassiness of localized electron systems was predicted long ago \cite{DaviesLee,Gruenewald82}. Experimentally, such behaviour was first found in the slow capacitance relaxation of uncompensated semiconductors~\cite{DonMonroe87}, and more recently in a series of experiments on strongly disordered indium-oxide films~\cite{Ovadyahu}, as well as in granular metallic films \cite{Grenet} where slow logarithmic relaxation, aging and memory were observed in the gate-controlled conductivity.

The memory of equilibration conditions is manifested as an anomalous (symmetric) field effect. The conductivity of the sample 
always increases no matter whether the carrier density is increased or decreased. This out-of-equilibrium dip in the 
conductivity 
is remarkably universal, its shape being independent of disorder and magnetic field, while its width (in the applied gate voltage) shrinks with decreasing temperature and increases with carrier density~\cite{Ovadyahu}. The latter fact demonstrates the importance of electron-electron interactions, suggesting that these systems constitute genuine electron glasses.

\section{Phenomenological theory of the memory effect}
In Ref.~\cite{LebanonMuller05}, we proposed a phenomenological theory to account for these features. 
The theory is based on the standard model of Anderson insulators (in 2D) with strongly localized, classical electrons~\cite{efrosshklovskii75}
\begin{equation}
\label{Hamiltonian}
H=\sum_{i\neq j} (n_i-\nu)\frac{e^2}{\kappa r_{ij}}(n_j-\nu) +\sum_i n_i \phi_i,
\end{equation}
the impurity sites being located on a square lattice with nearest neighbour distance $a$. Disorder and frustration are introduced by random site energies $\phi_i$, independently and uniformly distributed 
over the range 
$-W<\phi_i<W$, and $\nu$ is a homogeneously distributed background charge, adjustable by the gate.

Three key ingredients are essential
in order 
to understand the memory effect:

\textit{ i)} Below a critical temperature ($T<T_g$) the electron glass falls out of equilibrium on experimental time scales due to the appearance of a large number of metastable states which are separated by high barriers. After a sudden change of external parameters the system is unable to relax rapidly to its new ground state, and may retain a memory of its former state.
 This scenario is suggested by numerical simulations~\cite{Numerics}, as well as by mean field theory~\cite{MFTc} which predicts a glass 
transition at $T_g\sim (e^2/\kappa a)^2/W$ for large disorder $W\gg e^2/\kappa a$~\cite{note}. 

\textit {ii)} The repulsive interactions induce a Coulomb gap in the local density of states, which tends to the linear Efros-Shklovskii pseudogap at low temperatures~\cite{efrosshklovskii75}. 
The shape of the gap varies slightly from state to state and can serve as a fingerprint of the sample history.

\textit {iii)} The system conducts via variable range hopping which is exponentially sensitive to perturbations of the local density of states. 

The theory of Ref.~\cite{LebanonMuller05} makes the key assumption that the precise form of the Coulomb gap varies between different metastable states, and may thus reflect the pathway which a given state was obtained. 
This situation can occur when a small gate voltage is applied after a long equilibration. 
In the first place, the gate introduces additional background charges (changing $\nu\rightarrow \nu+\Delta \nu$ in (\ref{Hamiltonian})), forcing the same amount of charge to enter the system in order to compensate on average for the gate field. 
The new carriers occupy the empty states just above the Fermi level, $E_F$, shifting the Fermi level 
up with respect to the bulk of the occupied states. 
The resulting density of states is asymmetric around the new Fermi level.
We argue below that 
this feature survives the fast relaxations triggered by the presence of the new particles. 
Due to the asymmetry, thermal excitations of electrons into empty states will be enhanced, and the hopping conductivity will thus increase.
A quantitative analysis shows that this effect dominates over the decrease due to the suppression of hole excitations~\cite{LebanonMuller05}. 

\section{Metastability of out-of-equilibrium states}
The introduction of new carriers renders other particles unstable and induces spontaneous hops that decrease the total energy. The question arises whether, or to what extent, such fast relaxation processes can wipe out the initial asymmetry of the density of states.
In Ref.~\cite{LebanonMuller05}, we argued that as long as
the density of new carriers, $n_V$, is smaller than that of thermally excited electrons, $n_T$, the gate voltage
effect is perturbative. This condition is equivalent to the requirement that the shift of the Fermi level (with respect to the bulk of the states) be smaller than temperature. Below, we will provide more detailed arguments to justify the neglect of fast relaxations in this regime.
On the other hand, at higher gate voltages new carriers are introduced on sites
that were essentially always empty in the original state. Since in general their
local environment is not favorable to the addition of a particle, the new electrons likely trigger relaxation processes and destabilize the original state.
Such reconfigurations will lead to the saturation of the anomalous field effect at voltages where $n_V\geq n_T$.

Let us analyse the stability criterion $n_V<n_T$ in more detail in the case of low temperatures where the equilibrium density of states, $\rho_0(\xi\equiv E-E_F)$, displays a linear Coulomb gap. More precisely, we assume the scaling form $\rho_0(\xi)=T/(e^2/\kappa)^2 f(|\xi|/T)$ with $f(0)={\cal O}(1)$, and $f(x)\sim x$ for  $1\ll x\ll (e^2/\kappa a)^2/WT$. If the density of injected particles, $n_V$, is larger than $n_T\sim (\kappa T/e^2)^2$, their mutual interaction exceeds the temperature, so that they spontaneously rearrange, reshuffling the distribution of levels. On the other hand, for $n_V<n_T$, the interactions between new particles can be neglected. The particles will a priori occupy the first 
empty states above $E_F$, shifting the Fermi level by $\Delta E_F\approx \Delta \nu /\rho_0(0) \sim (e^2/\kappa)^2\Delta \nu/T$, so that the new density initially takes the form 
$ \rho_1(\xi)=\rho_0(\xi+\Delta E_F)$.
However, the Coulomb repulsion from the new electrons shifts the local energies of other sites, inducing spontaneous hops that affect the distribution of levels.   

Let us estimate the density $n_p(E^*)$ of recombining electron-hole pairs that affect the number of states in the interval $0<|\xi|<E^*$. We count the pairs ($ij$) whose recombination energy, $\Delta E_{ij}=e^2/\kappa r_{ij}-|\xi_i|-|\xi_j|$, is raised above temperature, 
while the pair was stable ($\Delta E_{ij}<T$) before the introduction of new particles. We approximate $n_p(E^*)$ by the density of states for pairs with $\Delta E_{ij}$ of order $T$, multiplied by the typical shift of $\Delta E_{ij}$ for a pair of sites with distance $r_{ij}$. The latter is of order 
 $r_{ij}\Delta \nu e^2/\kappa$ for the dominant short pairs with $r_{ij}<\Delta \nu^{-1/2}$. 
Assuming the single-site energies to be independent, we find
\begin{eqnarray}
\label{unstable}
n_p(E^*)\approx \int d^2r \int_0^{E^*} d\xi_1 \rho_1(\xi_1)\int_{-E^*}^0 d\xi_2 \rho_1(\xi_2)\, r \Delta \nu\frac{e^2}{\kappa}  \delta\left(\frac{e^2}{\kappa r} -\xi_1-|\xi_2|-T\right).
\end{eqnarray}
This should be compared to the change $n_d(E^*)$ in the density of particle/hole states in the same energy interval $|\xi|<E^*$, caused by the shift of $E_F$: $n_{d}(E^*)\approx \int_0^{E^*}(\rho_1(\xi)-\rho_0(\xi))d\xi \sim E^* \Delta E_F/(e^2/\kappa)^2$.
Performing the integration over $r$ in (\ref{unstable}), and dividing by $n_{d}(E^*)$, we obtain
\begin{eqnarray}
\frac{n_{p}(E^*)}{n_{d}(E^*)}\sim  \frac{\pi}{E^* \Delta E_F} \int_0^{E^*} d\xi_1 \rho_1(\xi_1)\int_{-E^*}^0 d\xi_2\rho_1(\xi_2)\frac{1}{(\xi_1+|\xi_2|+T)^3} \frac{e^4\Delta \nu}{\xi_1+|\xi_2|+T }\sim \frac{T}{E^*},
\end{eqnarray}
which is only a small fraction in the range of energies ($E^*\gg T$) that are probed in hopping conductivity. 
Note, however, that within the range $|\xi|\leq T$ the density of states can be appreciably modified by fast relaxations.

\section{Numerical simulations at $T=0$}
\begin{figure}
\includegraphics[width=0.5 \textwidth]{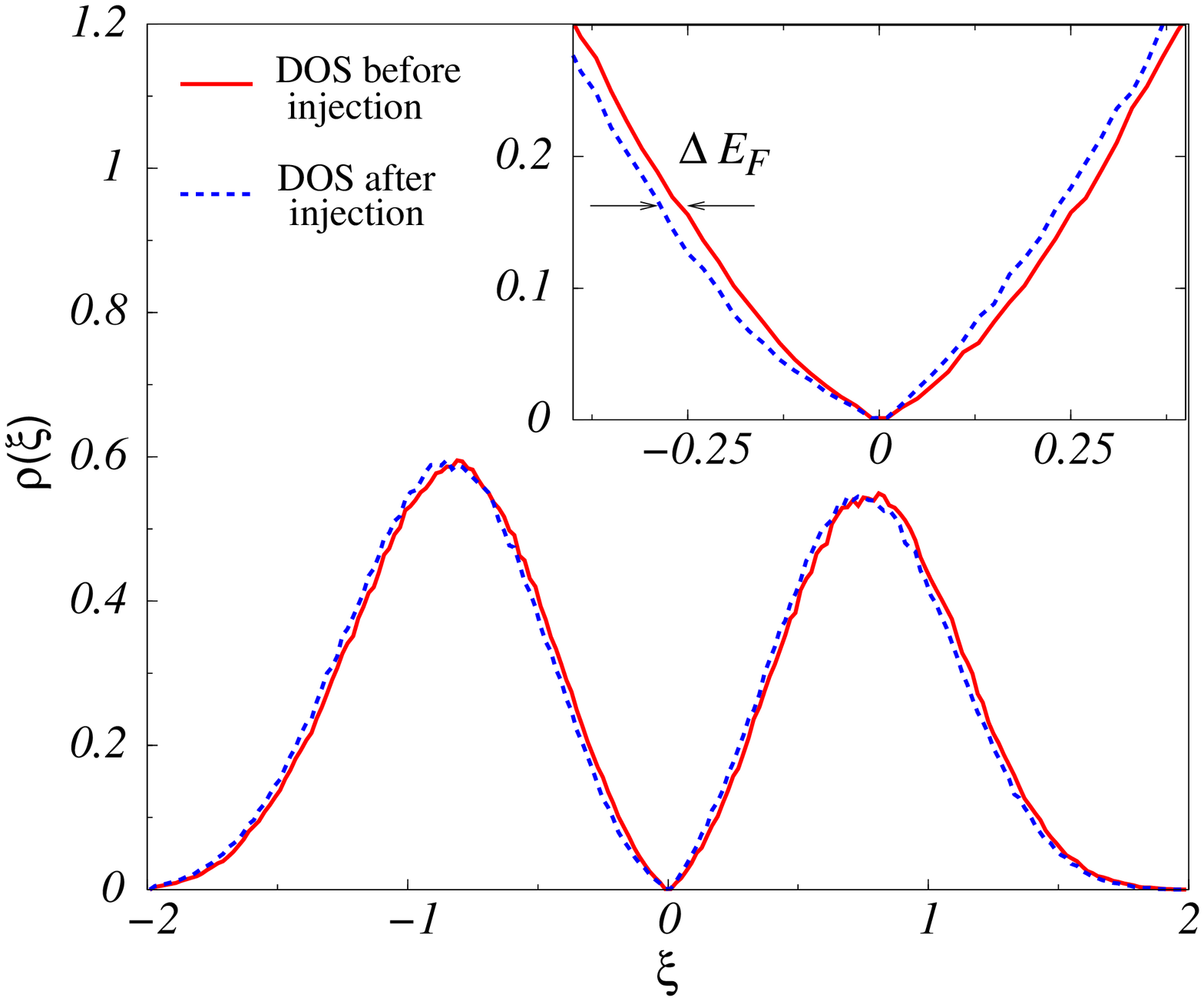}
\includegraphics[width=0.5 \textwidth]{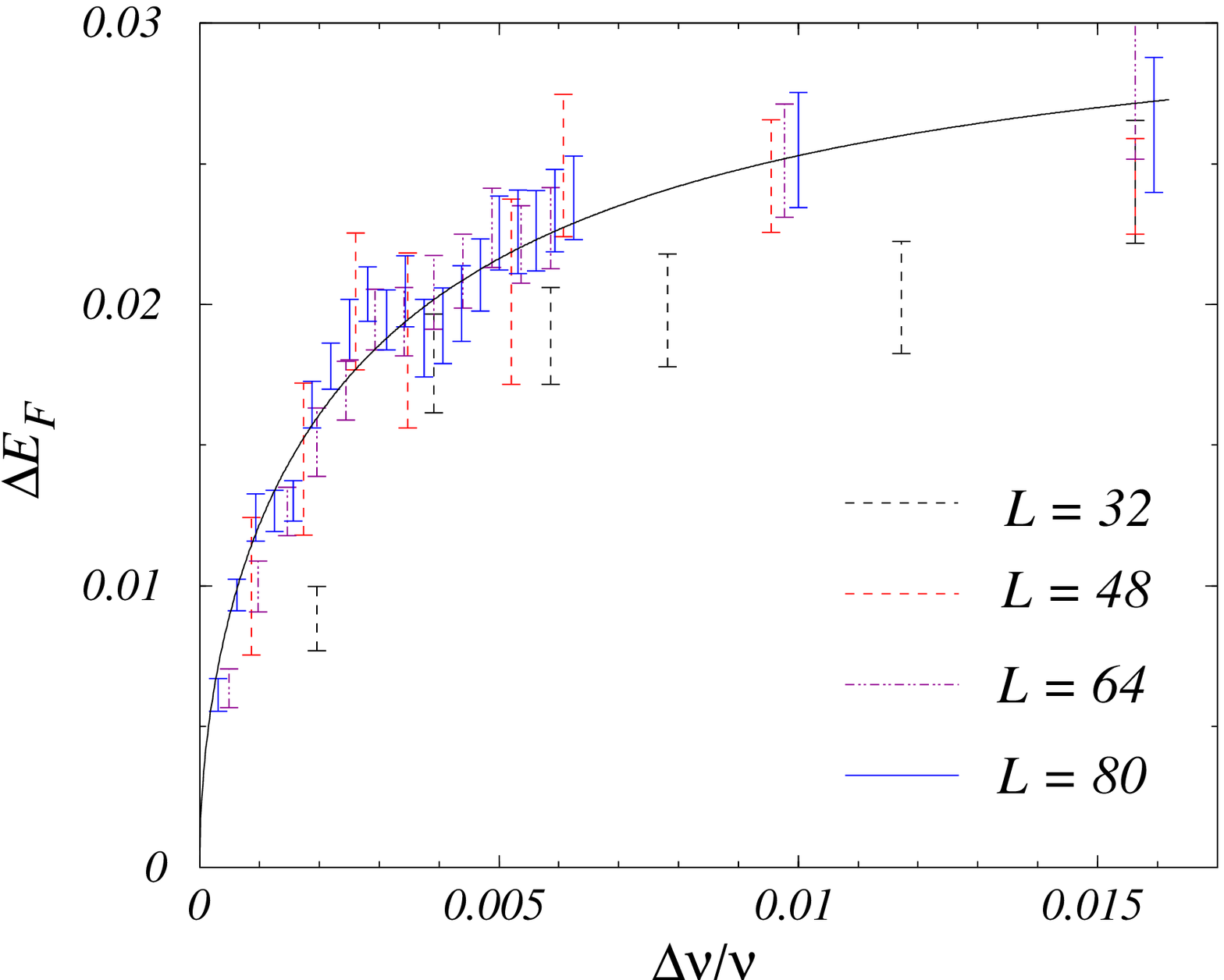}
\vspace{-5pt} \caption{
Left: The density of states before and after particle injection, plotted as a function of $\xi=E-E_F$. 
The data are averaged over 200 samples with 80$\times$80 sites, for disorder with width $W=0.6$. 
Right: The shift $\Delta E_F$ as a function of the relative increase of carrier density. $\Delta E_F$ is determined from the shift between the bulk of the density of states. It agrees within error bars with the difference between the average Fermi level in both kinds of states with the same filling fraction. 
The full line is the best fit to the form $\Delta E_F=a\sqrt{((\Delta\nu/\nu)^{-1}+c^{-1})^{-1}}$. 
}
\label{fig:shift}
\end{figure}

In order to provide evidence that history dependence can be seen in the density of states,
we carried out simulations at $T=0$ for the model (\ref{Hamiltonian})  with moderate disorder strength $W=0.6$ in systems of linear size $L=32,48,64,80$. Starting from random initial conditions, we allowed all hops that decrease the energy, until we reached a state that is stable to any single hop. We then filled in a fraction $\Delta \nu \ll \nu\approx 0.54$ of new particles, putting them onto the available sites with least energy cost. Finally, we simulated the fast relaxation processes by steepest descent via single hops to the closest metastable state. 
For both metastable states we determine the Fermi level $E_F$ (the energy separating occupied and empty levels) and calculate the distribution of local energies, $E_i=dH/dn_i$. For the states obtained before particle injection, $E_F$ is found to increase linearly with filling fraction, $d E_F/d\nu \approx 2.45$, as one expects for a system with finite (``field cooled'') compressibility.
However, in the states obtained after injection and relaxation, the average Fermi level is systematically higher than that of unbiased states with the same filling fraction. This is also manifested by the fact that the bulk of the density of states, $\rho_1(\xi)$, is shifted down by $\Delta E_F$, see Fig.~\ref{fig:shift}.
The extra shift $\Delta E_F$ increases with filling fraction as $\Delta\nu^{1/2}$, as one expects from the presence of a linear Coulomb gap. This result reflects the incompressibility of the Coulomb glass~\cite{DaviesLee}, $\kappa=(d\Delta E_F/d\Delta\nu)^{-1}\sim (\Delta \nu)^{1/2}\rightarrow 0$ (for small $\Delta \nu$). A similar phenomenon occurs in the pinned Wigner crystal~\cite{Giamarchi}.
At larger $\Delta \nu$, the shift saturates to $\Delta E_F^\infty$, as the system becomes unstable and reorganizes significantly. 

At finite, but small temperature the numerical results for $T=0$ should remain robust, since the density of states can only relax further by activated processes. 
The fact that at $T=0$ the shift $\Delta E_F$ saturates only at a finite fraction $\Delta \nu$ suggests that the saturation condition $n_V\approx n_T$ for the anomalous field effect only applies for $T>\Delta E_F^\infty$. However, at least in the presence of strong disorder, $W \gg e^2/\kappa a$, one expects a rather large temperature interval $\Delta E_F^\infty<T<T_g$ in which the width of the anomalous field effect is correctly described by $n_V\approx n_T$. 
 
In conclusion, we have provided quantitative arguments and numerical evidence at $T=0$ that the local density of states can exhibit memory of the sample history. These independent approaches provide further support for the memory scenario proposed in Ref.~\cite{LebanonMuller05}.  

MM was supported by NSF grant DMR 0210575. EL was supported by DOE grant DE-FE02-00ER45790.


\begin{thebibliography}{99}
\bibitem{note}
In low spatial dimensions (below the upper critical dimension) it is possible that there is no genuine thermodynamic transition. In this case, 
the mean field value for 
$T_g$ should be taken as an estimate of the temperature below which the dynamics is strongly activated.
\end{thebibliography}
\end{document}